\documentclass[letter,bibyear]{aa} 
\usepackage{natbib}
\usepackage{graphicx}
\usepackage{txfonts}

\newcommand{\jmst}{J.~Mol.~Struct.}   
\newcommand{\chemrev}{Chem.~Rev.}

\begin{document}

\title{Interstellar nitrile anions: Detection of C$_3$N$^-$ and C$_5$N$^-$ in TMC-1
\thanks{Based on observations carried out with the Yebes 40 m telescope (projects 19A003, 19A010, and
20A014) and the Institut de Radioastronomie Millim\'etrique (IRAM) 30 m telescope. 
  The 40 m radiotelescope at Yebes Observatory is operated by the Spanish Geographic Institute
  (IGN, Ministerio de Transportes, Movilidad y Agenda Urbana).
  IRAM is supported by INSU/CNRS (France), MPG (Germany), and IGN (Spain).}}

\author{
J.~Cernicharo\inst{1}\and
N.~Marcelino\inst{1}\and
J.~R~Pardo\inst{1}\and
M.~Ag\'undez\inst{1}\and
B.~Tercero\inst{2,3}\and
P.~de~Vicente\inst{2}\and
C.~Cabezas\inst{1}\and 
C.~Berm\'udez\inst{1} 
}

\institute{Grupo de Astrof\'isica Molecular, Instituto de F\'isica Fundamental (IFF-CSIC), C/ Serrano 121, 28006 Madrid, Spain.
email: jose.cernicharo@csic.es
\and Centro de Desarrollos Tecnol\'ogicos, Observatorio de Yebes (IGN), 19141 Yebes, Guadalajara, Spain.
\and Observatorio Astron\'omico Nacional (IGN), C/ Alfonso XII, 3, 28014, Madrid, Spain.
}

\date{Received; accepted}

\abstract{
We report on the first detection of C$_3$N$^-$ and C$_5$N$^-$ towards the cold dark core TMC-1 in the Taurus region, using the 
Yebes 40 m telescope.
The observed C$_3$N/C$_3$N$^-$ and C$_5$N/C$_5$N$^-$ abundance ratios are $\sim$140 and $\sim$2, respectively; that is similar
to those found in the circumstellar envelope of the carbon-rich star IRC\,+10216. Although the
formation mechanisms for the neutrals are different in interstellar (ion-neutral reactions) and circumstellar 
clouds (photodissociation and radical-neutral reactions), the
similarity of the C$_3$N/C$_3$N$^-$ and C$_5$N/C$_5$N$^-$ abundance ratios 
strongly suggests a common chemical path for the formation of these anions in interstellar and circumstellar clouds.
We discuss the role of radiative electronic
attachment, reactions between N atoms and carbon chain anions C$_n$$^-$, and that of H$^-$ reactions with HC$_3$N and HC$_5$N as possible routes to form C$_n$N$^-$.

The detection of C$_5$N$^-$ in TMC-1 gives strong support for assigning to this anion the lines found in IRC\,+10216, 
as it excludes the possibility of a metal-bearing species, or a vibrationally excited state.
New sets of rotational parameters have been derived from the observed frequencies in TMC-1 and IRC\,+10216
for C$_5$N$^-$ and the neutral radical C$_5$N.}


\titlerunning{C$_3$N$^-$ and C$_5$N$^-$ in TMC-1}
\authorrunning{Cernicharo et al.}

\maketitle

\section{Introduction}
The importance
of anions in the chemistry of interstellar clouds was analyzed in the early years
of astrochemistry  by \citet{Dalgarno1973}. The presence
of carbon chain negative ions in space was predicted on the ground that electron
radiative attachment is efficient for molecules with large electron
affinities and dense vibrational spectra \citep{Sarre1980,Herbst1981}. 

The first anion detected in space, C$_6$H$^-$, was observed towards TMC-1 \citep{McCarthy2006}.
Lines from this species were already reported as unidentified
features in the line survey of IRC\,+10216 performed with the Nobeyama 45 m telescope
by \citet{Kawaguchi1995}. Their assignation to C$_6$H$^-$ was not possible until
the laboratory observations of \citet{McCarthy2006} became available. Nevertheless, \citet{Aoki2000}
suggested that the carrier of these lines was C$_6$H$^-$ from ab initio calculations.
The laboratory and space detection of this species prompted attention 
to the abundance of hydrocarbon anions in interstellar and circumstellar clouds. C$_4$H$^-$
was first discovered in the circumstellar cloud IRC\,+10216 by \citet{Cernicharo2007}
and then in the interstellar clouds L1527 and TMC-1 \citep{Sakai2008,Agundez2008a,Cordiner2013}. 
C$_6$H$^-$ was
also detected towards other interstellar sources \citep{Sakai2007,Gupta2009,Cordiner2011,Cordiner2013}. 
Following the observation of C$_8$H$^-$ in the laboratory \citep{Gupta2007}
this anion was found in TMC-1 \citep{Brunken2007} and IRC\,+10216 \citep{Kawaguchi2007, Remijan2007}.

The nitrile anions CN$^-$, C$_3$N$^-$ and C$_5$N$^-$ were first detected in the circumstellar envelope of the carbon-rich 
star IRC\,+10216 \citep{Agundez2010,Thaddeus2008,Cernicharo2008}. While accurate laboratory
frequencies were available for
CN$^-$ and C$_3$N$^-$   \citep{Gottlieb2007,Thaddeus2008,Amano2008}, the assignment of C$_5$N$^-$
was based on the coincidence of the observed rotational constants with \emph{ab initio} calculations
by \citet{Botschwina2008} and \citet{Aoki2000}.
Although this species is the best candidate for this identification, the lack of precise laboratory frequencies
prevents us from ruling out other species involving metals, which are present in IRC\,+10216. For example,
MgC$_3$N and MgC$_4$H have been recently detected in IRC\,+10216 based on \emph{ab initio} calculations 
\citep{Cernicharo2019} and they have rotational constants $B$ just a few megahertz below that of C$_5$N$^-$.
Moreover, there is controversy about the formation of C$_n$N$^-$ anions through radiative 
electron attachment to C$_n$N radicals, for which calculated rate constants differ by orders of 
magnitude \citep{Walsh2009,Khamesian2016,Millar2017}. Hence, the detection of these anions in cold dark
clouds and the determination of their abundances
are an important step forward in understanding their chemistry in different astronomical environments.

In this Letter we present the detection of C$_3$N$^-$ and C$_5$N$^-$ in the cold dark core TMC-1. 
This is the first time nitrile anions are observed in the interstellar medium. 
The C$_3$N/C$_3$N$^-$ and C$_5$N/C$_5$N$^-$ abundance ratios derived are discussed within 
the frame of a chemical model of a cold dense cloud.

\begin{figure}[]
\centering
\includegraphics[scale=0.7,angle=0]{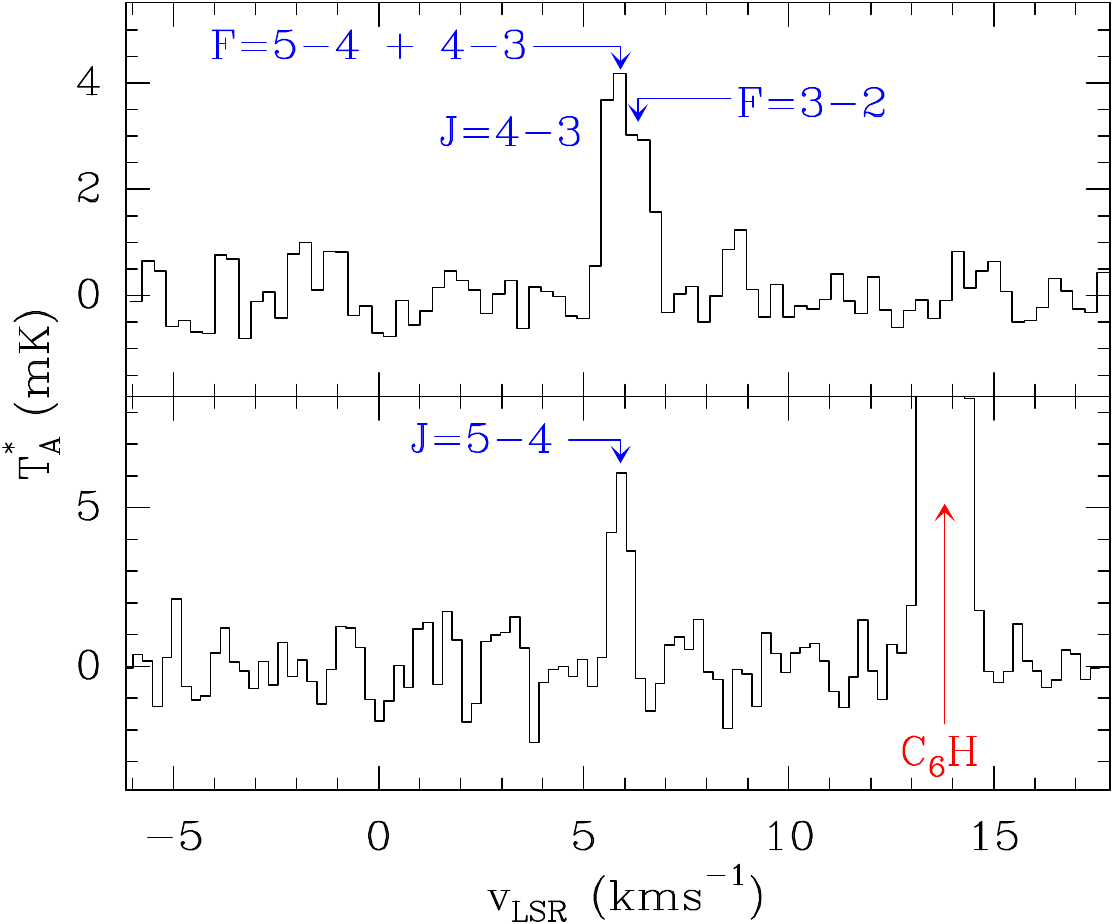}
\caption{Lines of C$_3$N$^-$ observed towards TMC-1 in the 31.0-50.3 GHz frequency range. The abscissa corresponds to the local
  standard of rest velocity in km s$^{-1}$. Frequencies and intensities for the observed lines are given in
  Table \ref{tab_c3nm_tmc1}. The ordinate axis represents the antenna temperature in mK corrected for atmospheric
  and telescope losses.}
\label{fig_c3nm_tmc1}
\end{figure}

\begin{figure}[]
\centering
\includegraphics[scale=0.6,angle=0]{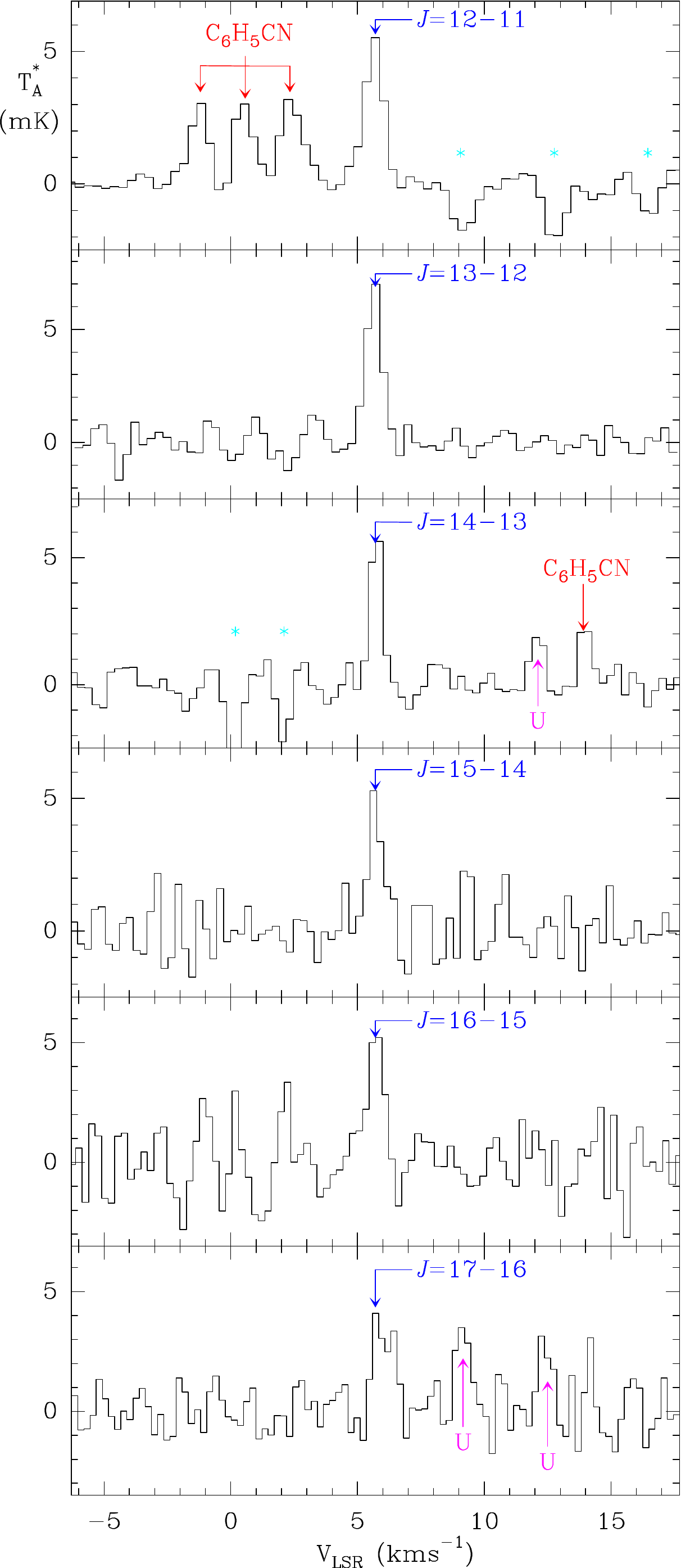}
\caption{Same as Fig. \ref{fig_c3nm_tmc1} but for C$_5$N$^-$. Observed frequencies and intensities are given in Table~\ref{tab_c5nm_tmc1}.}
\label{fig_c5nm_tmc1}
\end{figure}

\section{Observations} 

The Q-band (31.0-50.3 GHz) observations of TMC-1 ($\alpha_{2000}$=4$^h$ 41$^m$ 42.0$^s$,
$\delta_{2000}$=25$^o$ 41$'$ 27.6$''$) were carried out during the winter 2019/2020 with the 
40 m radio telescope of the Yebes observatory (IGN, Spain), hereafter Yebes 40 m. 
Observations of IRC\,+10216 in the same frequency band were performed in spring 2019 and have been 
previously described by \citet{Cernicharo2019} and \citet{Pardo2020}. New receivers were built within the Nanocosmos 
project\footnote{\texttt{https://nanocosmos.iff.csic.es/}} and installed at the telescope \citep{Tercero2020}. They were used 
for the observations presented in this work. The Q-band receiver consists of two HEMT cold amplifiers covering the 
31.0-50.3 GHz band with horizontal and vertical polarisations. Receiver temperatures 
vary from 22 K at 32 GHz to 42 K at 50 GHz. The backends are $16\times2.5$ GHz fast Fourier transform
spectrometers (FFTS) with a spectral resolution 
of 38.1 kHz providing the whole coverage of the Q band in both polarisations. 
The main beam efficiency varies from 0.6 at 32 GHz to 0.43 at 50 GHz. Pointing corrections
obtained by observing strong nearby quasars and were always within 2-3$''$.

For IRC\,+10216 the observing mode was position switching with an off position at 300$''$ in azimuth. 
The final spectra were smoothed to a resolution of 0.15 MHz, that is a velocity resolution 
of $\approx$1.5 and 0.9 km s$^{-1}$ at 31 and 50 GHz, respectively. The sensitivity of the final spectra 
varies between 0.4 mK and 1 mK per 0.15 MHz channel across the Q band, 
which is a factor of $\approx$10 better than previous observations 
in the same frequency range with the Nobeyama 45 m telescope taken with a spectral
resolution of 0.5-0.625 MHz \citep{Kawaguchi1995}. 

\begin{table}
\tiny
\caption{Observed line parameters for C$_3$N$^-$ in TMC-1} 
\label{tab_c3nm_tmc1}
\centering
\begin{tabular}{{cccccc}}
\hline \hline
Transition& $\nu_{rest}^a$& v$_{LSR}$   & $\Delta$v$^b$ & $\int$T$_A^*$dv $^c$ & T$_A^*$\\
          &     (MHz)     & (km s$^{-1}$)&  (km s$^{-1}$) & (mK km s$^{-1}$) & (mK)\\
\hline

{\textit J=4-3 F=4-3}     \\
{\textit    +  F=5-4}     & 38812.810 & 5.82$\pm$0.05 & 0.51$\pm$0.13& 2.4$\pm$0.3& 4.5\\
{\textit J=4-3 F=3-2}     & 38812.728 & 5.84$\pm$0.10 & 0.68$\pm$0.16& 2.3$\pm$0.4& 3.3\\
{\textit J=5-4      }     & 48515.874 & 5.90$\pm$0.03 & 0.50$\pm$0.07& 3.5$\pm$0.3& 6.5\\
\hline
\end{tabular}    
\tablefoot{\\
        \tablefoottext{a}{Rest frequencies as provided by the MADEX catalogue \citep{Cernicharo2012}. Typical
     uncertainties are 2 kHz.}\\
        \tablefoottext{b}{Line width at half intensity derived by fitting a Gaussian line profile to the observed
     transitions.}\\
        \tablefoottext{c}{Integrated line intensity. }\\
}                                                                                                                                                      
\end{table}
\normalsize

The TMC-1 Q-band observations were performed using the frequency switching technique with a 
frequency throw of 10\,MHz. The nominal spectral
resolution of 38.1 kHz was left unchanged for the final spectra
because of the low temperature of this source and, therefore, narrowness of its lines. The
average noise at this spectral resolution in the Q band ranges from $\sim$0.7 mK at 31 GHz
to $\sim$2 mK at 49.5 GHz,
which considerably improves previous line surveys in this frequency range for this source
\citep{Kaifu2004}. 

In order to improve the rotational constants of C$_5$N$^-$ we used all lines of this
species observed with the IRAM 30 m telescope since its detection \citep{Cernicharo2008}.
These observations in the $\lambda$ 3 mm band have been described in detail by \citet{Cernicharo2019}. They 
correspond to observations acquired during the last 35 years covering the 70-116 GHz range
with very high sensitivity (1-3 mK) over 1 MHz wide channels.
Examples of these data can be found in 
\cite{Cernicharo2004,Cernicharo2007,Cernicharo2008,Cernicharo2019} 
and \cite{Agundez2008b,Agundez2014}.

The beam size of the Yebes 40 m in the Q band is in the range 36-56$''$, while that of the IRAM 30 m telescope 
in the 3 mm domain is 21-30$''$. Pointing corrections were obtained by observing strong nearby quasars or
SiO masers for both sources. Pointing errors were always within 2-3$''$. 
The intensity scale for the observations with both telescopes, antenna temperature 
(T$_A^*$), was obtained after a calibration procedure that uses two absorbers at different
temperatures and the 
atmospheric transmission model (ATM; \citealt{Cernicharo1985, Pardo2001}). 
Calibration uncertainties are $\sim$10~\%. Additional uncertainties could arise, in
the case of IRC\,+10216, from the line intensity fluctuation induced by the time variation of the
stellar infrared flux \citep{Cernicharo2014,Pardo2018}. 
All data were analyzed via the GILDAS package\footnote{\texttt{http://www.iram.fr/IRAMFR/GILDAS}}. 

\begin{table}
\small
\caption{Observed line parameters for C$_5$N$^-$ in TMC-1} 
\label{tab_c5nm_tmc1}
\centering
\begin{tabular}{{ccrccc}}
\hline \hline
{\textit J$_u$}& $\nu_{obs}^a$& $\nu_{o}-\nu_{c}^b$  & $\Delta$v$^c$ & $\int$T$_A^*$dv $^d$ & T$_A^*$\\
               &  (MHz)       &     (kHz)            & (kms$^{-1}$)& (mK kms$^{-1}$)   & (mK)\\
\hline
12& 33332.572 &  1.7& 0.75$\pm$0.08& 5.0$\pm$0.2&  6.2\\
13& 36110.244 &  5.5& 0.73$\pm$0.07& 5.5$\pm$0.2&  7.1\\
14& 38887.893 & -2.6& 0.54$\pm$0.09& 3.7$\pm$0.3&  6.4\\
15& 41665.548 &  7.1& 0.50$\pm$0.16& 2.4$\pm$0.4&  4.6\\
16& 44443.178 &  4.5& 0.63$\pm$0.15& 3.3$\pm$0.5&  4.9\\
17& 47220.754 &-38.8& 0.76$\pm$0.19& 2.0$\pm$0.5&  2.5\\
\hline
\end{tabular}    
\tablefoot{\\
        \tablefoottext{a}{Observed frequencies for a v$_{LSR}$ of 5.83 km s$^{-1}$. The uncertainty is 10 kHz
    for all lines except for {\textit J=17-16,} for which it is 20 kHz.}\\
        \tablefoottext{b}{Observed minus calculated frequencies.}\\
        \tablefoottext{c}{Line width at half intensity derived by fitting a Gaussian line profile to the observed
     transitions.}\\
        \tablefoottext{d}{Integrated line intensity. }\\
}                                                                                                                                                      
\end{table}
\normalsize

\begin{figure}[]
\centering
\includegraphics[scale=0.3,angle=0]{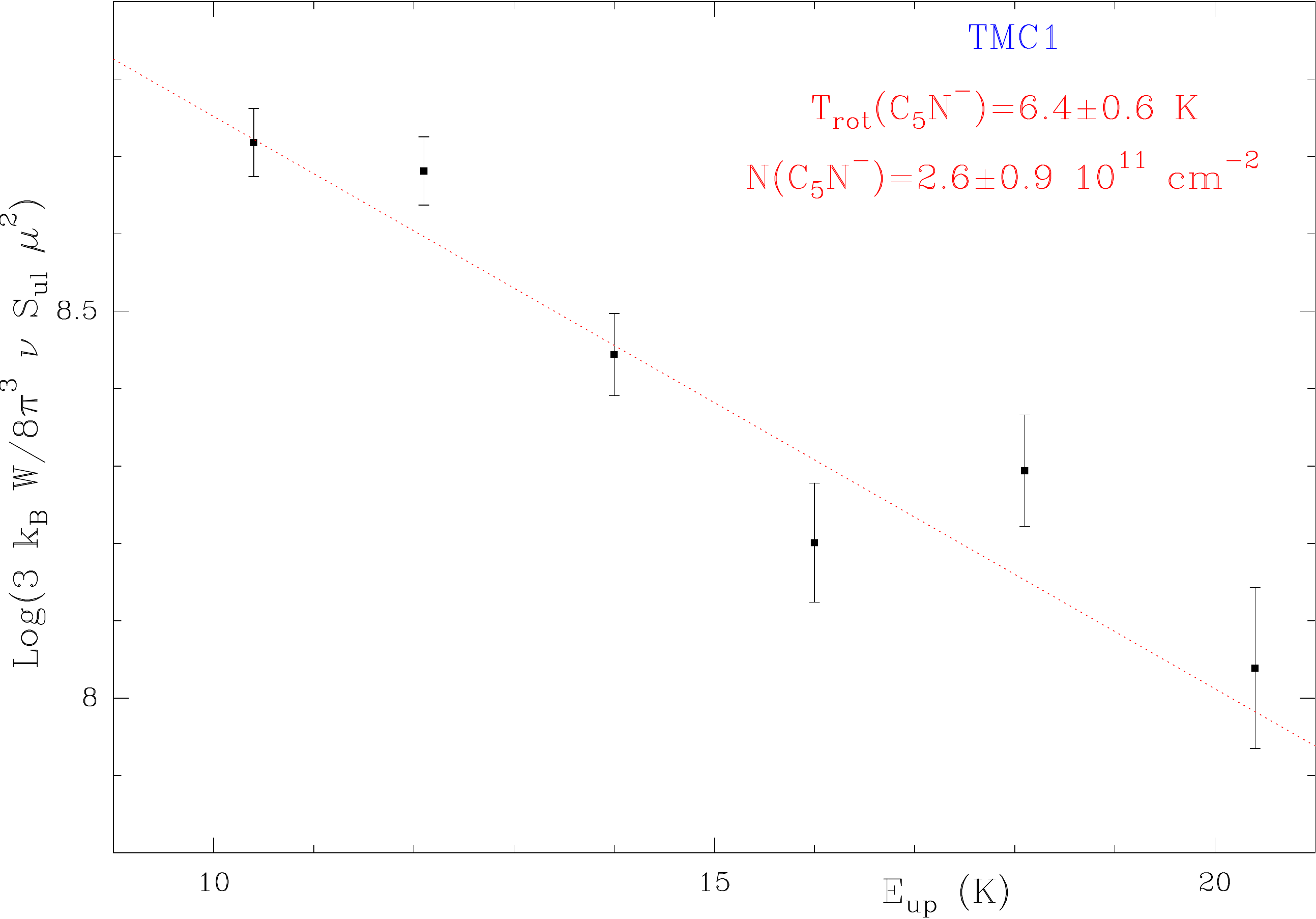}
\caption{Rotational diagram for the observed lines of C$_5$N$^-$ in TMC-1.}
\label{trot_tmc1}
\end{figure}

\section{Results} \label{sec:results}

One of the most remarkable results from the Q-band observations in TMC-1 and IRC\,+10216 is the presence of 
a forest of weak lines. Most of these can be assigned to known species and only a few remain unidentified
in IRC\,+10216 \citep{Cernicharo2019,Pardo2020} and TMC-1 (Marcelino et al., in preparation). For both
sources the level of sensitivity in this work has been increased, as commented previously, 
by a factor 5-10 with respect to previous works with other telescopes at the same frequencies 
\citep{Kawaguchi1995,Kaifu2004}. This high sensitivity 
has also allowed us to detect all known C$_{2n}$H and C$_{2n+1}$N anions in TMC-1. 
In this paper we focus 
on C$_3$N$^-$ and C$_5$N$^-$ lines in TMC-1.
In addition, we use the observed frequencies of C$_5$N$^-$ transitions towards IRC\,+10216 for an improved determination of the
rotational and distortion constants of this anion.

\subsection{C$_3$N$^-$}

Laboratory work exists for this anion
and frequencies are well determined with accuracies of $\simeq$2 kHz \citep{Thaddeus2008}. This species has two
rotational transitions in the 31.0-50.3 GHz frequency range.
Figure \ref{fig_c3nm_tmc1} shows both of these transitions towards TMC-1,  
while the derived line parameters are given in Table \ref{tab_c3nm_tmc1}. This species was
searched towards TMC-1 by \citet{Thaddeus2008}
without success, and to the best of our knowledge, this is the
first time this anion has been detected in an interstellar cloud.

\subsection{C$_5$N$^-$}
The detected lines of C$_5$N$^-$ towards TMC-1 are shown in Fig. \ref{fig_c5nm_tmc1}. Only the {\textit J}=18-17 transition 
at 49998.4 MHz has not been detected. This is, however, justified since the sensitivity of the data at 50 GHz
is 3 mK, whereas the expected {\textit J}=18-17 line intensity is below 5 mK (see  also Fig. \ref{fig_c5nm_tmc1}).
No C$_5$N$^-$ lines were detected with the Nobeyama 45 m telescope by \citet{Kaifu2004}. 

In order to derive precise frequencies for C$_5$N$^-$ in TMC-1 we
fitted, in our Yebes 40 m data, the central v$_{LSR}$ of the line emission 
from well-known species for which accurate laboratory rotational frequencies exist. From
the seven HC$_5$N rotational transitions in the 
Q band ({\textit J}$_{up}$=12 to {\textit J}$_{up}$=18), \citet{Cernicharo2020} 
derive a v$_{LSR}$ of 5.83$\pm$0.01 km s$^{-1}$. From the $^{13}$C and $^{15}$N isotopologues 
of HC$_5$N,
they obtain v$_{LSR}$ = 5.84$\pm$0.01 km s$^{-1}$. Hence, we
adopt a v$_{LSR}$ of 5.83 km s$^{-1}$ for further frequency determinations in TMC-1.
The value given by \citet{Kaifu2004} is 5.85 km s$^{-1}$, which is practically identical to our result within the
uncertainties. Derived line parameters for C$_5$N$^-$ in TMC-1 are given in Table \ref{tab_c5nm_tmc1}.

The characteristic 
U-shaped line profiles exhibited by molecular lines in IRC\,+10216 allow an accurate
central frequency determination, within $\simeq$50 kHz \citep{Cernicharo2018}, in
spite of the broad emission that covers 29 km s$^{-1}$ \citep{Cernicharo2000}. 
The v$_{LSR}$ of the source, -26.5 km s$^{-1}$, has been
well determined from the observation of hundreds of lines \citep{Cernicharo2000}. 
The observed  C$_5$N$^-$ lines in IRC\,+10216 at $\lambda$\,$\sim$\,7\,mm (Q band) are shown in Fig. \ref{fig_c5nm_irc}. Frequencies and
other line parameters are given in Table \ref{tab_c5nm_irc}. This table also includes lines observed in
the 3 mm data obtained after the detection of this species in 2008. All lines at 3 mm 
re-observed after \citet{Cernicharo2008} have a spectral
resolution of 0.198 MHz; these lines are shown in Fig. \ref{fig_irc_3mm}.

\subsection{New rotational constants for C$_5$N$^-$}
\label{new_rot_cons}
The frequencies of a linear molecular species can be fitted to this
standard expression involving the rotational quantum number J, rotational constant $B_0$,
and the distortion constant $D_0$:

$\nu(J\rightarrow$$J-1)$=2$B_0$J - 4$D_0$J$^3$.
\\
\noindent
Using all observed C$_5$N$^-$ lines in both sources, the fit provides the following results:

$B_0$=1388.86681(19) MHz

$D_0$=34.44(13)$\times$10$^{-6}$ MHz,

\noindent
where values between parentheses represent the 1$\sigma$ uncertainty for the fitted parameters.
The data were weighted in the fit according 1/$\delta v^2$, where $\delta v$ is the estimated uncertainty
on the observed frequencies.
The correlation coefficient between $B_0$ and $D_0$ is 0.738 and
the standard deviation between the predicted and observed frequencies is 66 kHz. These values significantly improve those derived from
previous observations by \citet{Cernicharo2008}, $B_0$=1388.860(2)
and $D_0$=33(1)$\times$10$^{-6}$ MHz.
We tried to fit the distortion constant of
order six ($H_0$), but the derived value is only marginally significant. Its inclusion reduces 
the weighted deviation from 1.03 to 0.89 and the standard deviation from 66 to 53.7 kHz. However, it
increases the correlation between the fitted parameters. The derived values in this case, are written as

$B_0$=1388.86734(25) MHz

$D_0$=35.68(44)$\times$10$^{-6}$ MHz

$H_0$=5.0(1.7)$\times$10$^{-10}$ MHz.
\\
The differences between observed and calculated frequencies are given in Tables \ref{tab_c5nm_tmc1}  
and \ref{tab_c5nm_irc} for TMC-1 and IRC+10216, respectively. New rotational parameters were also derived for C$_5$N (see Section \ref{app_c5n_rot}).

\section{Discussion}
Only two rotational C$_3$N$^-$ lines have been detected in TMC-1. We assumed a 
volume density of H$_2$ of 4$\times$10$^4$ cm$^{-3}$ (see e.g. \citealt{Fosse2001}),
a kinetic temperature of 10 K, a dipole moment for the molecule of 
2.27 D \citep{Pascoli1999}, 
and the collisional rates of C$_3$N$^-$/H$_2$ from \citet{Lara2019}. 
For simulating the source we assumed a circular uniform brightness distribution with
a radius of 40$''$ (see e.g. the intensity maps of different carbon chains presented
by \citealt{Fosse2001}). With this assumed size the source completely covers the main beam of the Yebes 40 m at 50\,GHz.

Using the large velocity gradient approximation (LVG) implemented in MADEX \citep{Cernicharo2012},
and correcting the intensities of all observed transitions
for the beam efficiency and the source beam dilution, we derive a column density
of {\textit N}(C$_3$N$^-$)=1.3$\times$10$^{11}$ cm$^{-2}$.

For C$_5$N$^-$ we have enough observed lines to build the rotational diagram shown in Fig \ref{trot_tmc1}. 
A dipole moment of 5.2 D was adopted for this species \citep{Botschwina2008}. A source size identical to that
of C$_3$N$^-$ was adopted. We derive
a rotational temperature of 6.4$\pm$0.6 K and a column density of (2.6$\pm$0.9)$\times$10$^{11}$ cm$^{-2}$.
We also performed an LVG calculation with MADEX adopting for C$_5$N$^-$ the collisional rates of 
C$_6$H$^-$/p-H$_2$ \citep{Walker2017}. For a kinetic temperature of 10 K, the observed intensities 
corrected for beam efficiency and
source beam dilution can be reproduced with a column 
density of 9$\times$10$^{11}$ cm$^{-2}$. If the collisional rates of HC$_5$N/p-H$_2$ are adopted
(F. Lique, private communication), then the column density is 1.5$\times$10$^{11}$ cm$^{-2}$. Hence,
the difference by a factor two of the latter value with respect to that from the rotational diagram is
due to the uncertainties on both the collisional rates and the assumed H$_2$ volume density. We adopt
for C$_5$N$^-$ the column density derived from the rotational diagram of Fig. \ref{trot_tmc1}.

The derived line parameters for C$_3$N and C$_5$N are given in Tables \ref{tab_c3n_tmc1} 
and \ref{tab_c5n_tmc1}.
For C$_3$N collisional rates are not available
and we assumed a rotational temperature of 7 K (that is similar to that of C$_5$N$^-$) to compute
line intensities assuming local thermodynamical equilibrium conditions (LTE).
Assuming the same source size as for the anions, we derive {\textit N}(C$_3$N)=1.8$\times$10$^{13}$ cm$^{-2}$.
For C$_5$N we can build a rotational
diagram based on the data provided in Table \ref{tab_c5n_tmc1}, which yields 
{\textit T$_{rot}$}=10.1$\pm$1.90 K and
{\textit N}(C$_5$N)=(6.0$\pm$2.5)$\times$10$^{11}$ cm$^{-2}$. 
The adopted dipole moment for this species is
3.385 D \citep{Botschwina1996}. Hence, the {\textit N}(C$_3$N)/{\textit N}(C$_3$N$^-$) and 
{\textit N}(C$_5$N)/{\textit N}(C$_5$N$^-$) abundance ratios in TMC-1 are $\sim$140, and 
$\sim$2.3, respectively.

In IRC\,+10216 \citet{Thaddeus2008} derived a N(C$_3$N)/N(C$_3$N$^-$) abundance ratio of $\simeq$194, and
\citet{Cernicharo2008} obtained  a N(C$_5$N)/N(C$_5$N$^-$) abundance ratio around 2.
Hence, the observed abundance ratios
between neutral radicals C$_n$N and their anions are very similar in interstellar and circumstellar
clouds.

One of the problems in obtaining the C$_n$N/C$_n$N$^-$ abundance ratio is the assumed permanent dipole moment
for the neutrals. In the case of C$_5$N, the dipole moment for the ground $^2\Sigma$ electronic state
was calculated by \citet{Botschwina1996}.
However, as discussed by \citet{Cernicharo2008}, the C$_5$N radical has a low lying $^2\Pi$ electronic state with a
dipole moment of $\sim$1 D \citep{Pauzat1991}. Detailed calculations by \citet{Botschwina1996}
indicate that it lies 500 cm$^{-1}$ above the $^2\Sigma$ ground state.
Hence, C$_5$N could have a dipole moment between these two values in the
case of admixing between the $^2\Sigma$ and the $^2\Pi$ states.
A dipole moment averaged over both electronic states (i.e., twice as small as
that calculated for the unperturbed $^2\Sigma$ state) would raise 
the C$_5$N/C$_5$N$^-$ abundance ratio to $\sim8$ in TMC-1 and IRC\,+10216, 
which is very similar to the C$_6$H/C$_6$H$^-$ ratio in IRC\,+10216 \citep{Cernicharo2007}.

The same situation applies to C$_4$H and C$_4$H$^-$. The neutral radical also has a $^2\Sigma^+$ ground electronic
state with a first excited $^2\Pi$ electronic state very close in energy. The case has been recently discussed
by \citet{Oyama2020} who computed a dipole moment for C$_4$H 2.4 times larger than the value of the
$^2\Sigma^+$ state alone. The effect is a decrease by a factor $\sim$6 on the derived C$_4$H column densities and
an increase of the C$_4$H/C$_4$H$^-$ abundance ratio by the same factor.

\begin{figure}[]
\centering
\includegraphics[width=\columnwidth,angle=0]{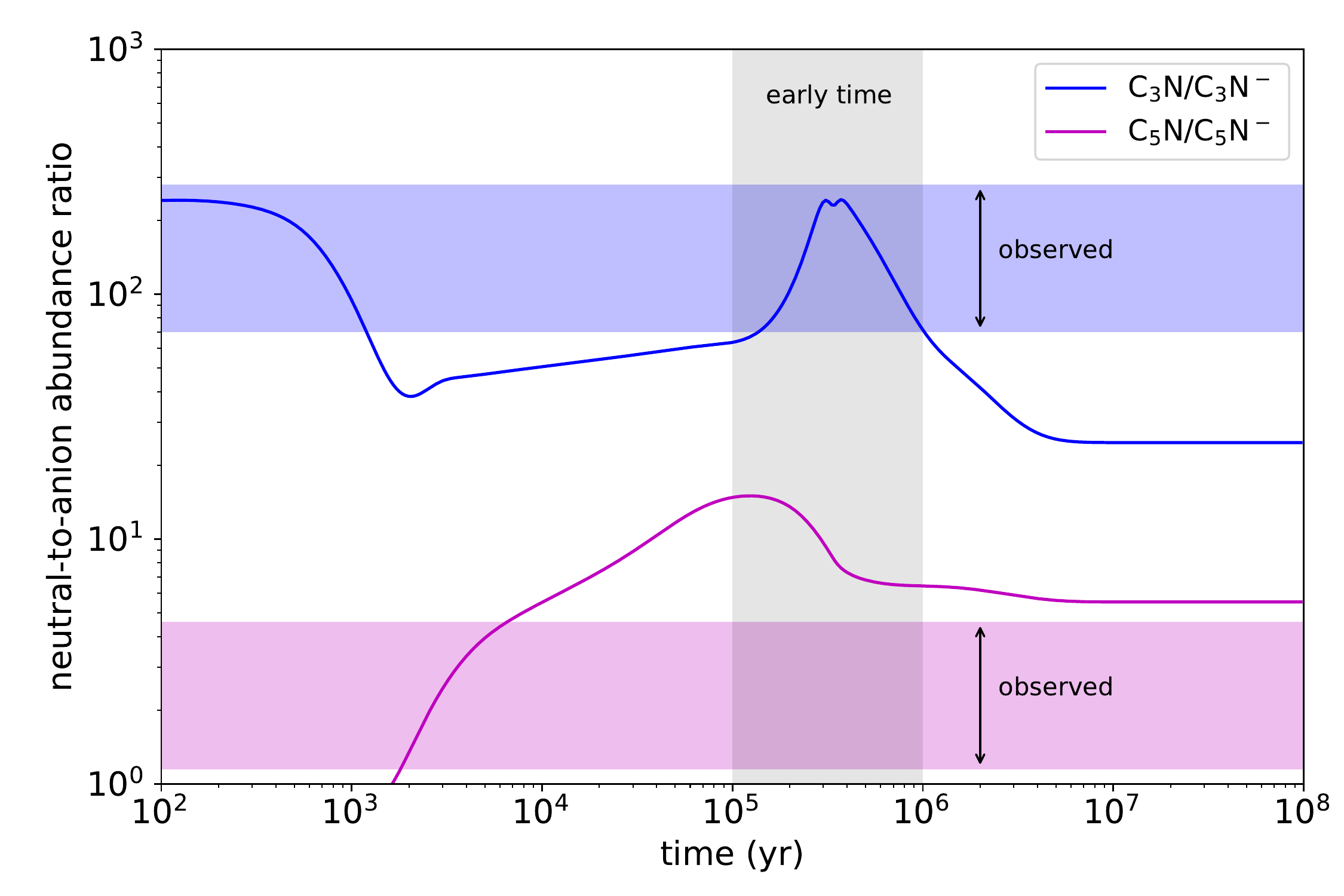}
\caption{Neutral-to-anion abundance ratios calculated with the chemical model are shown as a function of time and compared with observed values. The early time (10$^5$-10$^6$ yr) at which calculated abundances agree better with observations (see e.g. \citealt{Agundez2013}) is indicated. A conservative uncertainty of a factor of 2 is adopted for the observed ratios.}
\label{fig_ratios_anions}
\end{figure}

The chemistry of negative ions in cold interstellar clouds has been discussed 
by \cite{Walsh2009} and \cite{Millar2017}. In the light of the discovery of interstellar 
C$_3$N$^-$ and C$_5$N$^-$, we revisit the chemistry of these species in this work. We performed a 
chemical model of a cold dark cloud with typical physical parameters ($T_k$ = 10 K, 
$n_{\rm H}$ = $2\times10^4$ cm$^{-3}$, $\zeta$ = $1.3\times10^{-17}$ s$^{-1}$, $A_V$ = 
30 mag) and ``low metal'' elemental abundances (see \citealt{Agundez2013}). We adopted 
the University of Manchester Institute of Science and Technology
{\small RATE12} reaction network \citep{McElroy2013}, with a subset of reactions 
involving HCCN from \cite{Loison2015}. According to the model, formation of C$_3$N$^-$ 
does not occur via radiative electron attachment to C$_3$N, which is slow \citep{Petrie1997}, 
but through reactions between N atoms and bare carbon-chain anions C$_n^-$ (with $n \geq 6$), 
which are rapid and produce several nitrile radical and anions \citep{Eichelberger2007}. In 
the case of C$_5$N$^-$, the reaction of radiative electron attachment to C$_5$N is rapid 
\citep{Walsh2009} and dominates the synthesis of this anion. Formation of C$_3$N$^-$ through 
dissociative electron attachment to metastable isomers such as HNC$_3$ \citep{Harada2008} is 
included and occurs to some extent, although most of C$_3$N$^-$ is formed by N + C$_n$$^-$ 
reactions. Reactions of H$^-$ with HC$_3$N and HC$_5$N could be a source of C$_3$N$^-$ and 
C$_5$N$^-$, although they are likely to be too slow at 10 K based on calculations for similar 
reactions of H$^-$ with C$_2$H$_2$ and C$_4$H$_2$ \citep{Gianturco2016}. Destruction of 
C$_3$N$^-$ and C$_5$N$^-$ is dominated by reactions with neutral atoms H, O, and C in 
cold dark clouds. Nitrile anions have been shown to react rapidly with polar molecules 
at low temperatures \citep{Joalland2016}. However, it is unlikely that this is a major 
loss channel for anions in cold dense clouds because, apart from CO, molecules have 
much lower abundances than neutral atoms.

The much larger rate constant of radiative electron attachment 
to C$_5$N compared to C$_3$N is the main reason why C$_5$N$^-$ is calculated to be 
much more abundant with respect to the neutral than C$_3$N$^-$ (see Fig.~\ref{fig_ratios_anions}). 
The results of our model are similar to the theoretical predictions of \cite{Walsh2009}. 
Calculated neutral-to-anion abundance ratios during the so-called early time (10$^5$-10$^6$ yr), 
at which calculated abundances agree better with observations (see e.g. \citealt{Agundez2013}), 
agree well with the values retrieved from observations for C$_3$N$^-$, although they are 
somewhat higher for C$_5$N$^-$ (see Fig.~\ref{fig_ratios_anions}). We note that, as discussed 
above, the observed C$_5$N/C$_5$N$^-$ ratio could be as high as $\sim8$, in which case model and 
observations would agree much better. Finally, we point out that the chemical model predicts an 
abundance of $\sim3\times10^{-11}$ relative to H$_2$ for CN$^-$. Adopting a column density 
of 10$^{22}$ cm$^{-2}$ for H$_2$ \citep{Cernicharo1987}, the predicted column density of 
CN$^-$ would be $3\times10^{11}$ cm$^{-2}$, which is below the $3\sigma$ upper limit of 
$1.4\times10^{12}$ cm$^{-2}$ derived by \cite{Agundez2008a}.

This work definitively excludes 
metal-bearing species, or vibrationally excited states of other known species, as carriers for the 
series of lines assigned to C$_5$N$^-$ by \citet{Cernicharo2008}, and gives strong support to this
identification.

\begin{acknowledgements}

We thank Spanish Ministerio de Ciencia e Innovaci\'on for funding 
support through project AYA2016-75066-C2-1-P. We also thank ERC for funding through grant 
ERC-2013-Syg-610256-NANOCOSMOS. MA and CB thank Ministerio de Ciencia e Innovaci\'on  
for Ram\'on y Cajal grant RyC-2014-16277 and Juan de la Cierva grant FJCI-2016-27983. 
We thank support from French National Research  Agency through project Anion Cos Chem
(ANR-14-CE33-0013). We would like to thank our referee, Prof. S. Yamamoto, for useful
comments and suggestions.

\end{acknowledgements}

\normalsize

\begin{appendix}

\section{Additional tables and figures}
\begin{figure}[]
\centering
\includegraphics[scale=0.6,angle=0]{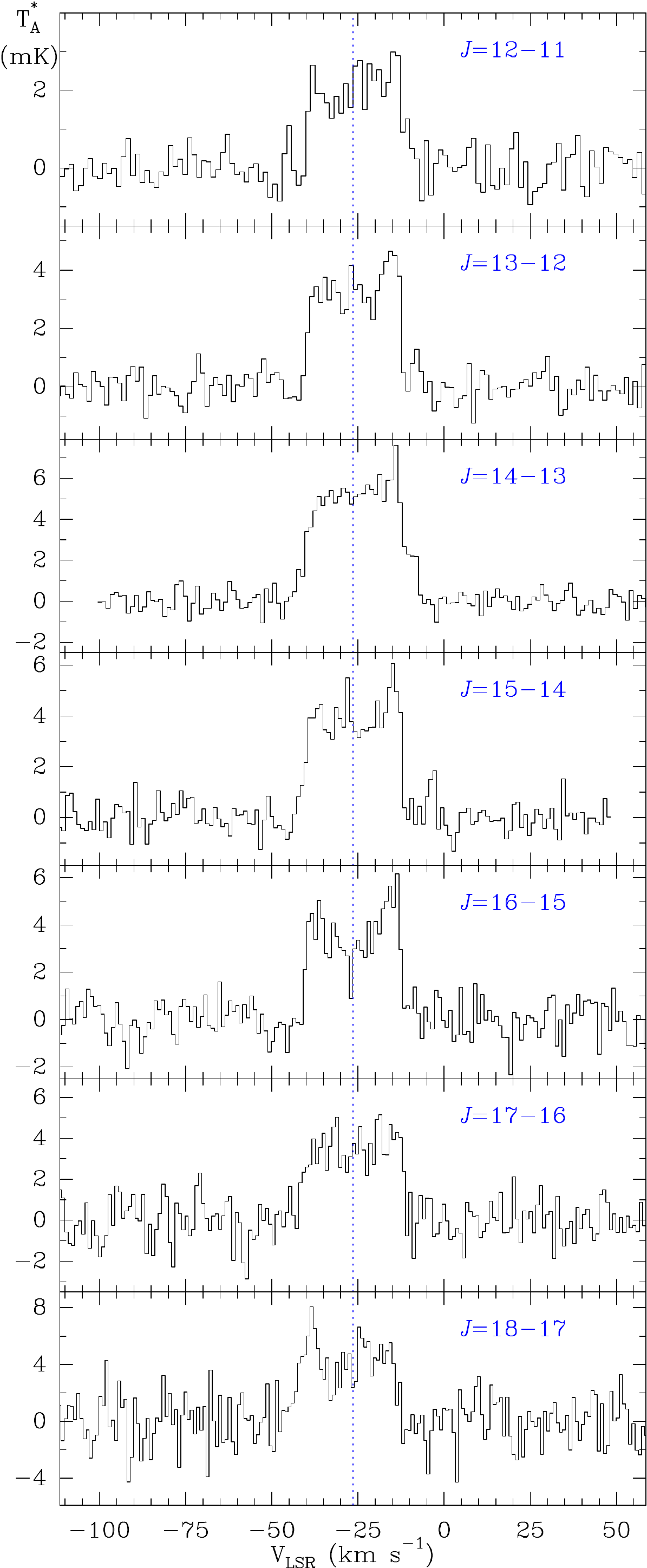}
\caption{Rotational transitions of C$_5$N$^-$ observed towards IRC\,+10216 in the 31-50 GHz range. Derived
    frequencies and intensities are given in Table \ref{tab_c5nm_irc}. Figure \ref{fig_irc_3mm} shows
    the lines of the same species observed with the IRAM 30 m telescope at $\lambda$=3mm .}
\label{fig_c5nm_irc}
\end{figure}

\begin{figure*}[]
\centering
\includegraphics[scale=0.82,angle=0]{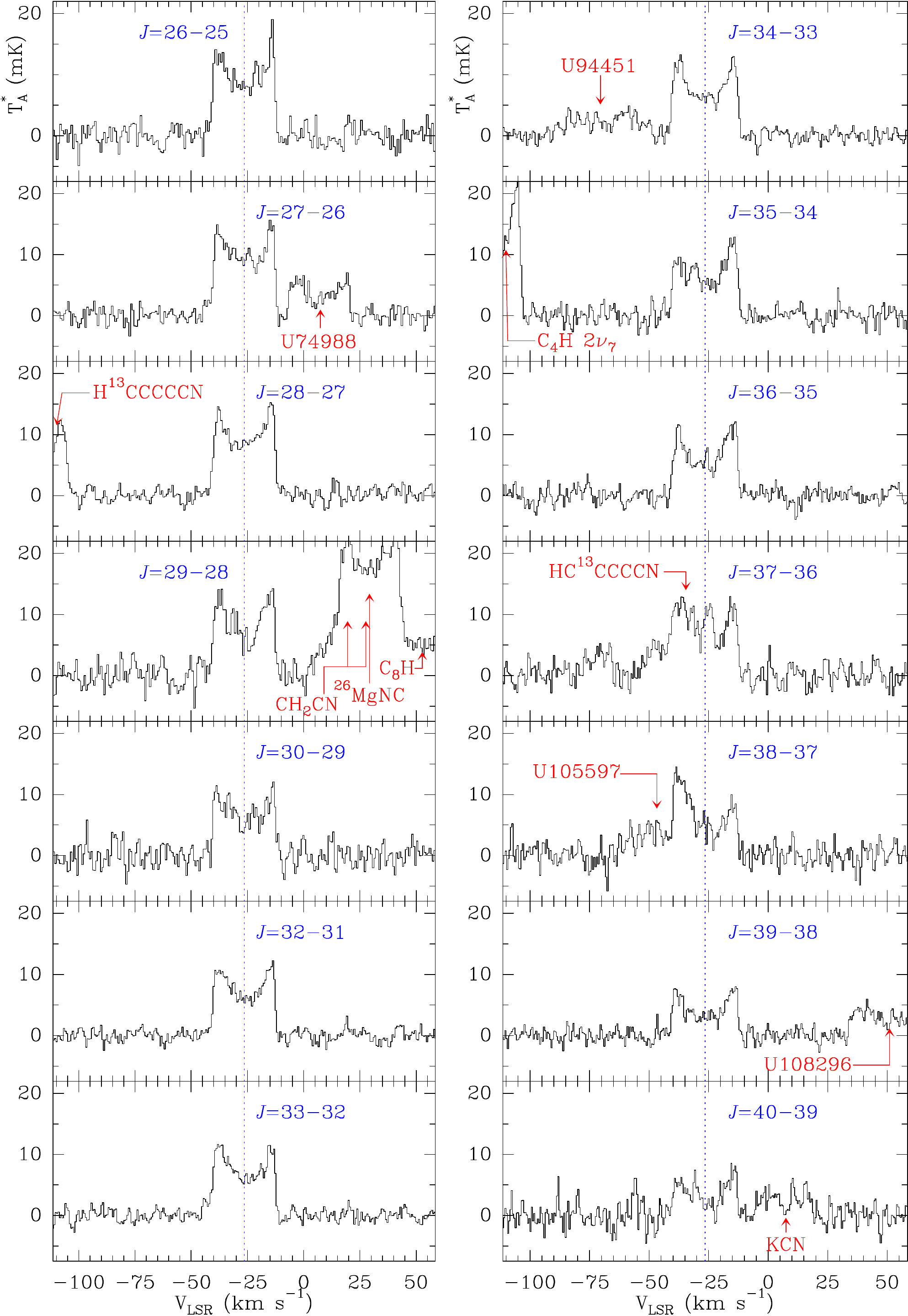}
\caption{Observed lines of C$_5$N$^-$ towards IRC\,+10216 in the 3\,mm domain with the IRAM 30 m telescope. 
Spectral resolution is 0.19 MHz for all lines. The  abscissa corresponds to the local standard of rest velocity
in km s$^{-1}$. Frequencies and intensities for the observed lines are given in Table \ref{tab_c5nm_irc}. 
The ordinate is the antenna temperature corrected for atmospheric and telescope losses in mK.}
\label{fig_irc_3mm}
\end{figure*}

Line parameters for C$_5$N$^-$ towards IRC\,+10216 are given in Table \ref{tab_c5nm_irc}. The 
observed lines are shown in
Figures \ref{fig_c5nm_irc} and \ref{fig_irc_3mm}. 
The data in the 3 mm domain (see Fig. \ref{fig_irc_3mm}) considerably improve the signal to noise ratio of the lines reported by \citet{Cernicharo2008}. The new
rotational constants derived from these frequencies and those of TMC-1 are discussed in Section
\ref{new_rot_cons}.

The observed line parameters for C$_3$N and
C$_5$N in TMC-1, which are needed to compute the column density of these species and to compare with that of
their anions, are given in Tables \ref{tab_c3n_tmc1} and \ref{tab_c5n_tmc1}.
For C$_3$N, accurate
frequency predictions can be found in the Cologne Database for Molecular Spectroscopy (CDMS; \citealt{Muller2005}). 
However, we discuss below the
case of C$_5$N, as significant deviations between predicted and measured frequencies in space are observed.

\begin{table}
\caption{Observed line parameters for C$_5$N$^-$ in IRC\,+10216} 
\label{tab_c5nm_irc}
\centering
\begin{tabular}{{ccccc}}
\hline \hline
{\textit J$_u$} & $\nu_{obs}^a$ & $\nu_{o}-\nu_{c}^b$ &$\int$Tdv$^c$& Notes\\
   &   (MHz)       &      (MHz)            & (K km s$^{-1}$)&\\
\hline
12&  33332.508$\pm$0.050&-0.062 & 0.063$\pm$0.003&  \\ %
13&  36110.197$\pm$0.050&-0.040 & 0.091$\pm$0.003&  \\ %
14&  38887.789$\pm$0.100&-0.105 & 0.154$\pm$0.005& 1\\ 
15&  41665.527$\pm$0.050&-0.013 & 0.115$\pm$0.003&  \\ %
16&  44443.116$\pm$0.050&-0.057 & 0.108$\pm$0.003&  \\ %
17&  47220.740$\pm$0.050&-0.052 & 0.102$\pm$0.003&  \\ %

18&  49998.450$\pm$0.100&+0.052 & 0.100$\pm$0.005& 1\\ 
26&  72218.580$\pm$0.050&-0.049 & 0.303$\pm$0.008&  \\ %
27&  74996.039$\pm$0.050&-0.032 & 0.308$\pm$0.005&  \\ 
28&  77773.492$\pm$0.050&+0.002 & 0.297$\pm$0.004&  \\ %
29&  80550.817$\pm$0.050&-0.070 & 0.264$\pm$0.009&  \\ %
30&  83328.258$\pm$0.050&-0.003 & 0.216$\pm$0.007&  \\ %
31&                     &       &                & 2\\ 
32&  88882.883$\pm$0.050&-0.052 & 0.233$\pm$0.003&  \\ %
33&  91660.297$\pm$0.050&+0.063 & 0.232$\pm$0.003&  \\ %
34&  94437.518$\pm$0.050&+0.011 & 0.238$\pm$0.003&  \\ %
35&  97214.796$\pm$0.050&+0.042 & 0.216$\pm$0.006&  \\ %
36&  99991.999$\pm$0.050&+0.026 & 0.207$\pm$0.006&  \\ %
37& 102769.266$\pm$0.100&+0.101 & 0.266$\pm$0.007& 3\\ 
38& 105546.319$\pm$0.050&-0.008 & 0.202$\pm$0.006&  \\ %
39& 108323.470$\pm$0.050&+0.010 & 0.130$\pm$0.003&  \\ %
40& 111100.565$\pm$0.100&+0.002 & 0.111$\pm$0.008&  \\ %
41& 113877.556$\pm$0.100&-0.081 & 0.092$\pm$0.015& 1\\ 
48& 133316.175$\pm$0.150&-0.078 & 0.056$\pm$0.010& 4\\ 
\hline
\end{tabular}    
\tablefoot{\\
        \tablefoottext{a}{Observed frequencies assuming a source v$_{LSR}$ of -26.5 kms$^{-1}$ and an expanding
     velocity of 14.5 kms$^{-1}$ \citep{Cernicharo2000,Cernicharo2018}.}\\
        \tablefoottext{b}{Observed minus calculated frequencies.}\\
        \tablefoottext{c}{Integrated line intensity in K kms$^{-1}$.}\\
        \tablefoottext{1}{Blended with another feature but clear line profile.}\\
        \tablefoottext{2}{Fully blended with another spectral feature. Fit not possible.}\\
        \tablefoottext{3}{Data smoothed to a resolution of 0.6 MHz.}\\
        \tablefoottext{4}{Data smoothed to a resolution of 1.0 MHz.}\\
}                                                                                                                                                      
\end{table}
 
\begin{table*}
\caption{Observed line parameters for C$_3$N in TMC-1} 
\label{tab_c3n_tmc1}
\centering
\begin{tabular}{{cccccc}}
\hline \hline
{\textit N$_u$ J$_u$ F$_u$ $\rightarrow$ N$_l$ J$_l$ F$_l$ } 
&                $\nu_{rest}^a$     &  v$_{LSR}$ & $\int$Tdv$^b$  & $\Delta$v$^c$ &  T$_a^*$  \\
                &                    (MHz)          &(km s$^{-1}$)&(mK kms$^{-1}$)& (kms$^{-1}$)& (mK) \\
\hline
4 4.5 4.5$\rightarrow$3 3.5 4.5& 39571.110$\pm$0.027 &5.78$\pm$0.02&   5.7$\pm$1.0& 0.61$\pm$0.05&   8.8\\
4 4.5 3.5$\rightarrow$3 3.5 2.5& 39571.319$\pm$0.003 &5.72$\pm$0.01& 136.7$\pm$1.0& 0.56$\pm$0.01& 230.6\\
4 4.5 4.5$\rightarrow$3 3.5 3.5& 39571.326$\pm$0.003$^f$ &         &            &              &      \\
4 4.5 5.5$\rightarrow$3 3.5 4.5& 39571.397$\pm$0.003 &5.73$\pm$0.01& 160.0$\pm$3.0& 0.51$\pm$0.02& 297.2\\
4 4.5 3.5$\rightarrow$3 3.5 3.5& 39574.032$\pm$0.044 &5.61$\pm$0.02&   6.0$\pm$1.0& 0.53$\pm$0.04&  10.7\\
4 3.5 3.5$\rightarrow$3 2.5 3.5& 39587.553$\pm$0.048 &5.86$\pm$0.03&   7.8$\pm$1.0& 0.59$\pm$0.04&  12.3\\
4 3.5 2.5$\rightarrow$3 2.5 1.5& 39590.129$\pm$0.004 &5.73$\pm$0.01&  45.4$\pm$1.0& 0.36$\pm$0.01& 119.3\\
4 3.5 3.5$\rightarrow$3 2.5 2.5& 39590.204$\pm$0.004 &5.70$\pm$0.01& 179.5$\pm$3.0& 0.56$\pm$0.01& 301.2\\
4 3.5 4.5$\rightarrow$3 2.5 3.5& 39590.212$\pm$0.004$^f$ &         &            &              &      \\
4 3.5 2.5$\rightarrow$3 2.5 2.5& 39591.004$\pm$0.028 &5.62$\pm$0.02&   7.6$\pm$1.0& 0.58$\pm$0.05&  12.4\\
5 5.5    $\rightarrow$4 4.5$^e$& 49466.424$\pm$0.003 &5.70$\pm$0.02& 201.4$\pm$3.0& 0.66$\pm$0.04& 286.9\\
5 4.5    $\rightarrow$4 3.5$^e$& 49485.227$\pm$0.005 &5.67$\pm$0.02& 164.8$\pm$3.0& 0.64$\pm$0.04& 242.1\\
\hline
\end{tabular}    
\tablefoot{\\
        \tablefoottext{a}{Rest frequency from the CDMS catalogue \citep{Muller2005}.}\\
        \tablefoottext{b}{Integrated line intensity in K kms$^{-1}$.}\\
        \tablefoottext{c}{Line width at half intensity derived by fitting a Gaussian line profile to the observed
     transitions (in kms$^{-1}$).}\\
        \tablefoottext{d}{Integrated line intensity in mK kms$^{-1}$.}\\
        \tablefoottext{e}{Unresolved hyperfine structure. Rest frequencies correspond
     to the weighted value of the three stronger components.}\\
        \tablefoottext{f}{Two hyperfine components are unresolved. Rest frequencies correspond
     to the weighted value of these two components.}\\
    
}                                                                                                                                                      
\end{table*}
\begin{table}
\caption{Observed line parameters for C$_5$N in TMC-1} 
\label{tab_c5n_tmc1}
\centering
\begin{tabular}{{cccccc}}
\hline \hline
{\textit N$_u$ J$_u$} & $\nu_{obs}^a$ & $\int$Tdv$^b$  & $\Delta$v$^c$ &  T$_a^*$  \\
                      &    (MHz)      &(mK kms$^{-1}$)& (kms$^{-1}$)& (mK) \\
\hline
12 12.5& 33668.248& 7.4$\pm$0.5&   0.80$\pm$0.04& 8.8\\
12 11.5& 33678.986& 6.6$\pm$0.5&   0.85$\pm$0.05& 7.4\\
13 13.5& 36474.328& 6.3$\pm$0.7&   0.65$\pm$0.31& 9.1\\
13 12.5& 36485.068& 4.3$\pm$0.6&   0.52$\pm$0.05& 7.7\\
14 14.5& 39280.400& 5.8$\pm$0.6&   0.55$\pm$0.05& 9.9\\
14 13.5& 39291.136& 5.1$\pm$0.8&   0.62$\pm$0.07& 7.8\\
15 15.5& 42086.452& 3.6$\pm$0.9&   0.44$\pm$0.07& 7.8\\
15 14.5& 42097.179& 2.9$\pm$1.0&   0.46$\pm$0.08& 6.0\\
16 16.5& 44892.479& 3.7$\pm$1.0&   0.64$\pm$0.14& 5.4\\
16 15.5& 44903.223& 3.5$\pm$1.0&   0.39$\pm$0.10& 8.4\\
17 17.5& 47698.500& 4.1$\pm$1.1&   0.48$\pm$0.13& 8.2\\
17 16.5& 47709.233& 7.5$\pm$1.1&   0.78$\pm$0.12& 9.0\\
\hline
\end{tabular}    
\tablefoot{\\
        \tablefoottext{a}{Observed frequency in MHz assuming a v$_{LSR}$ of 5.83 km s$^{-1}$. Frequency uncertainty is 10 kHz
     for all lines.}\\
        \tablefoottext{b}{Integrated line intensity in K kms$^{-1}$.}\\
        \tablefoottext{c}{Line width at half intensity derived by fitting a Gaussian line profile to the observed
     transitions (in kms$^{-1}$).}\\
    }                                                                                                                                                      
\end{table}

\subsection{New rotational parameters for C$_5$N}
\label{app_c5n_rot}
For C$_5$N the frequencies of rotational lines up to $J$=6 ($\nu_{max}$=16.842 GHz)
were measured in the laboratory by \citet{Kasai1997}. These authors
provide frequency predictions up to  98.2 GHz that are systematically below the observed frequencies 
of this species towards TMC-1 (data provided in Table \ref{tab_c5n_tmc1})
and IRC\,+10216 (data provided in Table \ref{tab_c5n_irc}). In order to derive more accurate predictions
we fitted the observed astronomical frequencies to the same Hamiltonian than \citet{Kasai1997}. 
The fits
only consider frequencies measured in astronomical (space) sources that have an unresolved hyperfine structure
and a merged set of
space and laboratory frequencies.
The results are provided in Table \ref{tab_c5n_rot}. The poorly determined distortion constant from the laboratory
data alone is responsible for the observed discrepancies. The new $D$ value derived from space data alone is 44.2$\pm$0.2 Hz, compared with
the laboratory value of 50$\pm$10 Hz. 
In IRC\,+10216 several additional doublets of C$_5$N are detected
between $J_{up}$=40 to $J_{up}$=48. However, their line intensity is weak and the derived frequencies have uncertainties
of $\sim$1 MHz. These have not been included in the fit. 

In addition we combined in one fit all the rotational transitions 
measured so far for C$_5$N, including those from laboratory measurements and those observed in 
TMC-1 and IRC\,+10216. We used the SPFIT program \citep{Pickett1991}. The 
results are shown in Table \ref{tab_c5n_rot}. 
In all these fits the data were weighted according 1/$\sigma^2$, where $\sigma$ is the estimated uncertainty
on the measured frequencies.  
We obtained new values for the rotational and 
centrifugal distortion constants for the three spin-rotation constants and also for the nuclear quadruple coupling 
constant. In total we analyzed 69 rotational transitions. We tried to determine additional parameters such as the distortion 
constant $H$ or the spin-rotation constant $\gamma$$_D$, but the attempts resulted to be unfruitful. The uncertainty in the
rotational constant, $B$, improves from 0.54 kHz at \citet{Kasai1997} to 0.15 kHz. The most significant change corresponds
to the distortion constant, which has an uncertainty of 10 Hz from the laboratory data alone, and 0.3 Hz when the 
lines from TMC-1 and IRC\,+10216 are included.

The fit to the TMC-1 and IRC\,+10216 alone allows us to predict frequencies without hyperfine splitting with an accuracy
better than 0.1 MHz 
up to $N_{up}$=55 ($\nu\sim$154.3 GHz).
Observable hyperfine splitting in TMC-1 could occur for transitions with $N\leq$8 ($\nu\sim$22.4 GHz). In this case, the merged fit to
the laboratory and space data is recommended to compute the expected frequencies.

\begin{table}
\caption{Observed line parameters for C$_5$N in IRC\,+10216} 
\label{tab_c5n_irc}
\centering
\begin{tabular}{{cccc}}
\hline \hline
{\textit N$_u$ J$_u$} & $\nu_{obs}^a$ & $\int$T$_A^*$dv$^b$  & Notes\\
                      &    (MHz)      &(mK kms$^{-1}$)      &      \\
\hline
26 26.5&  72951.721$\pm$0.050& 18$\pm$6& 1,A \\
26 25.5&  .........          & ....... & 2   \\
27 27.5&  75757.571$\pm$0.050& 23$\pm$3& A   \\
27 26.5&  75768.214$\pm$0.100& 29$\pm$5& 1,B \\
28 28.5&  78563.437$\pm$0.100& 18$\pm$3& B   \\
28 27.5&  78573.910$\pm$0.100& 19$\pm$3& B   \\
29 29.5&  81368.856$\pm$0.150& 32$\pm$2& C   \\
29 28.5&  81379.846$\pm$0.150& 28$\pm$2& C   \\
30 30.5&  84174.795$\pm$0.200& 31$\pm$6& C   \\
30 29.5&  .........          & ....... & 2   \\
31 31.5&  86980.324$\pm$0.150& 25$\pm$3& C   \\
31 30.5&  .........          & ....... & 2,C \\
32 32.5&  89785.932$\pm$0.150& 27$\pm$1& C   \\
32 31.5&  89796.811$\pm$0.150& 33$\pm$2& C   \\
33 33.5&  92591.638$\pm$0.150& 24$\pm$1& C   \\
33 32.5&  92602.409$\pm$0.150& 29$\pm$1& C   \\
34 34.5&  95397.144$\pm$0.200& 43$\pm$7& 1,C \\
34 33.5&  .........          & ....... & 2   \\
35 35.5&  98202.488$\pm$0.150& 24$\pm$3& C   \\
35 34.5&  98213.325$\pm$0.200& 15$\pm$5& 1,C \\
36 36.5& 101008.060$\pm$0.150& 30$\pm$3& C   \\
36 35.5& 101018.872$\pm$0.150& 26$\pm$3& C   \\
37 37.5& 103813.438$\pm$0.150& 32$\pm$4& C   \\
37 36.5& 103824.263$\pm$0.150& 32$\pm$4& 1,C \\
38 38.5& 106618.712$\pm$0.200& 46$\pm$4& C   \\
38 37.5& 106629.939$\pm$0.200& 29$\pm$4& C   \\
39 39.5& ..........          & ....... & 2   \\
39 38.5& ..........          & ....... & 2   \\
40 40.5& 112229.929$\pm$0.500& 25$\pm$5& C   \\
40 39.5& 112240.834$\pm$0.500& 28$\pm$6& C   \\
\hline
\end{tabular}    
\tablefoot{\\
        \tablefoottext{a}{Observed frequency assuming a v$_{LSR}$ of -26.5 km s$^{-1}$ \citep{Cernicharo2000,Cernicharo2018}.}\\
        \tablefoottext{b}{Integrated line intensity in mK km s$^{-1}$. Expanding velocity fixed to 14.5 km s$^{-1}$
     \citep{Cernicharo2000,Cernicharo2018}.}\\
        \tablefoottext{1}{Partially blended with another spectral feature. Fit still possible.}\\
        \tablefoottext{2}{Fully blended with another stronger feature. Unreliable fit.}\\
        \tablefoottext{A}{Spectral resolution of 0.2 MHz.}\\
        \tablefoottext{B}{Spectral resolution of 0.6 MHz.}\\
        \tablefoottext{C}{Spectral resolution of 1.0 MHz.}\\ 
}                                                                                                                                                      
\end{table}

\begin{table*}
\caption{New derived rotational parameters for C$_5$N from TMC-1 \& IRC\,+10216} 
\label{tab_c5n_rot}
\centering
\begin{tabular}{{|c|c|c|c|c|c|}}
\hline
Constant$^a$  & Lab$^b$       &   TMC-1$^c$     &  TMC-1 \& IRC\,+10216 $^d$ & Space + Lab$^e$\\
\hline
$B$           & 1403.07981(54)& 1403.07995(19) & 1403.08041(13)& 1403.07974(15)\\
$D$           & 0.000050(10)  & 0.00004324(41) & 0.00004423(20)& 0.00004354(30)\\
$\gamma$      &-10.7472(35)   &-10.7363(23)    &-10.7360(44)   &-10.7478(27) \\
$b$           &  1.583(13)    & ........       & ........      &  1.5810(94 )\\
$c$           & -3.613(25)    & ........       & ........      & -3.615(18)  \\
$eQq$         & -4.341(17)    & ........       & ........      & -4.342(13)  \\
$N_{max}^f$   &    6          &   17           &  40           &    40       \\
$\sigma$(kHz)$^g$&   10       &  4.1           & 122           &    96       \\
\hline
\end{tabular}    
\tablefoot{\\
        \tablefoottext{a}{Rotational constants in MHz.}\\
        \tablefoottext{b}{Laboratory rotational parameters from \citet{Kasai1997}.}\\
        \tablefoottext{c}{Rotational parameters derived from the observed line frequencies in TMC-1 (see Table \ref{tab_c5n_tmc1}).}\\
        \tablefoottext{d}{Rotational parameters derived from the observed line frequencies in TMC-1 (see Table \ref{tab_c5n_tmc1}) 
     and  IRC\,+10216 (see Table \ref{tab_c5n_irc}).}\\
        \tablefoottext{e}{Merged fit to the laboratory and the TMC-1 and IRC\,+10216 data.}\\ 
        \tablefoottext{f}{ Highest rotational quantum number observed.}\\ 
        \tablefoottext{g}{Standard deviation of the fit in kHz.}\\ 
}                                                                                                                                                      
\end{table*}

\end{appendix}

\end{document}